\documentclass{ifacconf}

\makeatletter
\let\old@ssect\@ssect
\makeatother

\usepackage{enumerate}
\usepackage{amsmath}
\usepackage{amssymb}
\usepackage{graphicx}
\usepackage{mathtools}
\usepackage{booktabs}
\usepackage{xcolor, soul}
\usepackage{siunitx}    
\usepackage{lipsum}
\usepackage{mathabx}
\usepackage{accents}
\usepackage{comment}
\usepackage{tablefootnote}
\usepackage{makecell}
\usepackage[caption=false,font=footnotesize]{subfig}
\usepackage{algorithm}
\usepackage{algpseudocode}
\usepackage{natbib} 
\usepackage{hyperref}
\usepackage{subfloat}
\usepackage[firstpageonly=true]{draftwatermark}	

\DeclareMathOperator{\cat}{cat}
\DeclareMathOperator{\uncat}{uncat}
\DeclareMathOperator{\diag}{diag}
\DeclareMathOperator{\conjop}{conj}
\DeclareMathOperator{\FFT}{FFT}
\DeclareMathOperator{\iFFT}{iFFT}
\DeclareMathOperator{\MSE}{MSE}

\newcommand\scalemath[2]{\scalebox{#1}{\mbox{\ensuremath{\displaystyle #2}}}}

\newcommand{\ubar}[1]{\underaccent{\bar}{#1}}

\makeatletter
\def\@ssect#1#2#3#4#5#6{%
  \NR@gettitle{#6}
  \old@ssect{#1}{#2}{#3}{#4}{#5}{#6}
}
\makeatother

\hypersetup{
	colorlinks,
	linkcolor={red!40!black},
	citecolor={blue!40!black},
	urlcolor={blue!80!black}
}

\begin{document}
\begin{frontmatter}

\title{Structured state-space models are \newline deep Wiener models} 
 \thanks[footnoteinfo]{This research was financially supported by \emph{Kjell och M{\"a}rta Beijer Foundation} and by the project \emph{Deep probabilistic regression -- new models and learning algorithms} (contract number: 2021-04301), funded by the Swedish Research Council.}

\author[uu]{Fabio Bonassi}, 
\author[uu]{Carl Andersson}, 
\author[uu]{Per Mattsson},
\author[uu]{Thomas B. Schön}

\address[uu]{Department of Information Technology, Uppsala University,\\75105 Uppsala, Sweden. E-mail: {\tt name.surname@it.uu.se}}

\begin{abstract}    
The goal of this paper is to provide a system identification-friendly introduction to the Structured State-space Models (SSMs).
These models have become recently popular in the machine learning community since, owing to their parallelizability, they can be efficiently and scalably trained to tackle extremely long sequence classification and regression problems.
Interestingly, SSMs appear as an effective way to learn deep Wiener models, which allows us to reframe SSMs as an extension of a model class commonly used in system identification.
To stimulate a fruitful exchange of ideas between the machine learning and system identification communities, we deem it  useful to summarize the recent contributions on the topic in a structured and accessible form.
At last, we highlight future research directions for which this community could provide impactful contributions.
\end{abstract}


\end{frontmatter}

 \DraftwatermarkOptions{%
 angle=0,
 hpos=0.5\paperwidth,
 vpos=0.97\paperheight,
 fontsize=0.012\paperwidth,
 color={[gray]{0.2}},
 text={
   \parbox{0.99\textwidth}{\copyright 2024 the authors. This work has been accepted to IFAC for publication under a Creative Commons Licence CC-BY-NC-ND.}},
 }
 
\section{Introduction}
Recent years have been characterized by a remarkable research interest towards deep learning tools for data-driven control. 
An interesting use case is that of nonlinear system identification. 
Over the years, a variety of (increasingly complex) Neural Network (NN) architectures have indeed been proposed for nonlinear system identification, ranging from gated Recurrent NNs (RNNs) \citep{bonassi2022survey} to Transformer NNs (TNNs)  \citep{sun2022efficient}.
These models have proven to work well in many challenging identification problems, enabling the synthesis of accurate model predictive control laws, e.g. \cite{lanzetti2019recurrent}. 

Despite their modeling power, these architectures are known to suffer from computational efficiency problems at training. 
On the one hand, RNNs are inherently sequential models, implying they have to be iteratively unrolled over the time axis $T$ of training sequences \citep{bianchi2017recurrent}.
On the other hand, TNNs are plagued by a quadratic scaling issue ($\mathcal{O}(T^2)$), thus calling for tailored architectures like Longformers, which are however still untapped in the nonlinear system identification realm.
Unfortunately, these problems have limited the use of RNNs and TNNs for long-term nonlinear system identification to those cases where enough computational budget is available. 

What is more, both RNNs and TNNs are still fairly unexplored when it comes to their system-theoretical properties, such as Incremental Input-to-State Stability ($\delta$ISS) \citep{bonassi2022survey}, which are however essential for utilizing these models for control design \citep{bonassi2024nonlinear}.
Most of the research in this area is tailored to specific RNNs, see \cite{miller2018stable, bonassi2022survey}, and mainly involves stability promotion via regularization.

To address these problems, \cite{gu2021efficiently} proposed a \textbf{Structured State-space Model (SSM)} architecture named S4, which consists of multiple layers composed by LTI discrete-time systems followed by a nonlinear function. 
The term ``structured'' stems from the fact that this LTI system is given a specific structure to improve the architecture's modeling performances while also reducing the computational cost at training \citep{yu2018identification}. 
Nonlinear state-space models are not new, see \cite{marconato2013improved}, yet their adoption has been hampered by their crucial reliance on the model structure and on the initialization method of learnable parameters.
The contribution of the S4 approach towards SSMs has therefore been that of providing (\emph{i}) a novel, intrinsically stable, parametrization of the LTI system obtained by discretizing a continuous-time Diagonal Plus-Low Rank (DPLR) system, (\emph{ii}) a new strategy towards the parameters' initialization problem, (\emph{iii}) a computationally efficient approach to simulate (and train) these models over extremely long sequences, and (\emph{iv}) an empirical proof of the state-of-the-art performances of these models in long-term sequence learning problems.

Motivated by these appealing features, many works have continued to build on the S4 architecture.
    For example, \cite{gupta2022diagonal} and \cite{gu2022parameterization}  have explored the benefits entailed by stricter SSM structures, namely the parametrization via diagonal continuous-time systems (S4D), and by simpler initialization strategies.
\cite{smith2022simplified} have explored a novel, and somewhat more computationally efficient, simulation method for diagonal continuous-time parametrizations, named S5.
\cite{orvieto2023resurrecting} recently investigated the parametrization of the LTI subsystems directly in the discrete time domain, resulting in the Linear Recurrent Unit (LRU) architecture.

\subsubsection{Contribution} 
Despite the appealing results achieved by SSMs in the long-range arena benchmarks sequence classification problems, their use for nonlinear system identification is still unexplored. 
With this paper, we want to change that by making the following contributions. 
First of all we show that it is possible to interpret SSMs as \emph{deep Wiener models}, i.e. model structures where several Wiener models are interconnected in series. 
An interesting note here is that even though the Wiener models have been extremely popular within system identification --- see e.g. \cite{schoukens2017identification} and references therein --- their structure has been limited to ``single-layer'' {or parallel architectures \citep{wills2012generalised}}. 
Our second contribution is to dissect the recent developments on SSMs and explain them in terms of their structure and parameterization, and to clearly separate this from their initialization, simulation, and training strategies. 
The presentation in the paper is also done using the language commonly used in the system identification community in order to speed up the use of these tools within this area.

\subsubsection{Notation}
The imaginary unit is denoted by $i = \sqrt{-1}$.
Given a vector $v$, we denote by $v^\prime$ its real transpose.
For a time-dependent vector, the discrete-time index $k$ is reported as a subscript, e.g., $v_k$. 
Moreover, we denote by $v_{a:b}$ (where $a \leq b$) the sequence $v_{a:b} = \big( v_{a}, v_{a+1}, ..., v_{b} \big)$.
For this sequence, we indicate by $\cat(v_{a:b})$ the concatenation of its elements, i.e. $\cat(v_{a:b}) = [ v_a^\prime,  ..., v_{b}^\prime ]^\prime$, {and by $\uncat(\cdot)$ its inverse operation returning a sequence of vectors given their concatenation}.
Given a complex matrix $A$, we let $\conjop(A)$ be its element-wise complex conjugate and $A^*$ be its Hermitian transpose. 
Diagonal matrices may be defined via the $\diag$ operator, as $A = \diag(a_1, ..., a_n)$.

\section{Structured State-space Models}

\begin{figure}[t]
    \centering    \includegraphics[width=0.9\linewidth]{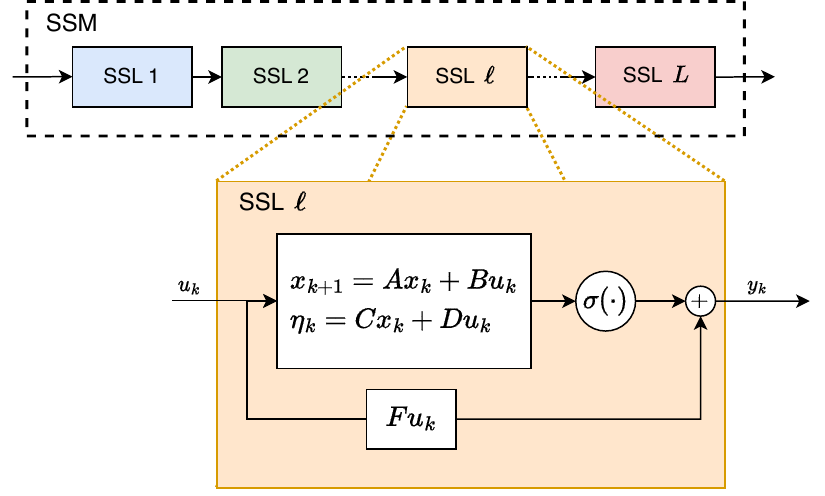}
    \caption{Schematic of a Structured State-space Model.}
    \label{fig:deepwiener}
\end{figure}

Consider the model depicted in Figure \ref{fig:deepwiener}, which consists of $L$ Wiener systems interconnected in series.
Each of these layers is here referred to as \textbf{Structured State-space Layer (SSL)}. 
Their interconnection results in an SSM, which can be interpreted as a specific configuration of a deep Wiener system.
We let the generic $\ell$-th SSL  ($\ell \in \{1, ..., L\}$) be represented by a discrete-time state-space model
\begin{equation} \label{eq:ssl:discrete}
    \text{SSL}_{\ell}: \,\begin{dcases}
        x_{k+1} = A x_k + B u_k, \\
        \eta_{k} = C x_k + D u_k, \\
        y_k = \sigma(\eta_k ) + F u_k, 
    \end{dcases}
\end{equation} 
where, for compactness, the layer index is omitted.
System~\eqref{eq:ssl:discrete} is characterized by the input vector $u \in \mathbb{R}^{n_u}$, the intermediate vector $\eta \in \mathbb{R}^{n_y}$, the output vector $y \in \mathbb{R}^{n_y}$, and the complex-valued state vector $x \in \mathbb{C}^{n_x}$.   
The SSL is parametrized by the matrices $\{ A, B, C, D, F \}$. 
The output transformation $\sigma(\cdot)$ can be any nonlinear, Lipschitz-continuous activation function, such as the $\tanh$, ELU, or Swish, see \cite{ramachandran2017searching}. 
In what follows, we aim to provide an overview of the possible structure, parametrization, initialization, and simulation strategies for this SSL. 

\smallskip
\begin{rem} \label{remark:deep}
    When a deep SSM is considered ($L > 1$), each layer is parametrized and initialized independently from the others. The simulation is carried out iteratively over the set of layers, meaning that the output sequence of the $\ell$-th layer is used as input of the layer $\ell+1$.
\end{rem}

We start by observing that, because both the input and output vector of the SSL are real-valued, the system matrices $A$, $B$, and $C$ of \eqref{eq:ssl:discrete} might be re-written as
\begin{equation} \label{eq:ssl:conj}
\begin{aligned}
    A =& \begin{bmatrix}
        \tilde{A} & \\ & \conjop(\tilde{A})
    \end{bmatrix}, \quad B = \begin{bmatrix}
        \tilde{B} \\ \conjop(\tilde{B})
    \end{bmatrix}, \\
     C =& \begin{bmatrix}
            \tilde{C} & \conjop(\tilde{C})
        \end{bmatrix},
\end{aligned}
\end{equation}
where $\tilde{A} \in \mathbb{C}^{n_\lambda \times n_\lambda}$, $\tilde{B} \in \mathbb{C}^{n_\lambda \times n_u}$, $\tilde{C} \in \mathbb{C}^{n_y \times n_\lambda}$, while $D \in \mathbb{R}^{n_y \times n_u}$, with $n_\lambda = \frac{n_x}{2}$.
Matrix $F \in \mathbb{R}^{n_y \times n_u}$ can be interpreted as a skip connection, {fixed to $I$ if $n_u = n_y$.
Note that if the SSLs are parametrized to be structurally Schur stable, the stability of the SSM is guaranteed.}

\vspace{0.5mm}
\begin{prop} \label{prop:diss}
    {If each layer $\ell \in \{1, ..., L \}$ is parametrized by a Schur-stable matrix $\tilde{A}$, then the SSM is Incrementally Input-to-State Stable ($\delta$ISS) \cite[Def.~4]{bonassi2022survey}.}
\end{prop}

\vspace{-3mm}
\begin{pf}
    {
    Owing to the Schur stability of $\tilde{A}$ and Lipschitz continuity of $\sigma(\cdot)$, the SSL can be easily proven (globally) $\delta$ISS.
    The SSM is then the series of $\delta$ISS systems and is therefore $\delta$ISS \citep{angeli2002lyapunov}. $\hfill\qed$}
\end{pf}

{While} \eqref{eq:ssl:conj} ensures that the eigenvalues come in complex-conjugate pairs,  both in the linear \citep{yu2018identification} and the nonlinear realm \citep{marconato2013improved} it is well known that state-space models call for additional structure to reduce the number of learnable parameters and hence achieve improved modeling performances and computational efficiency.
{Besides, the initialization of these parameters is also paramount to achieving satisfactory performances.
In the following, the strategies proposed in recent literature to address these problems are discussed.}

\section{Discrete-time SSL parametrizations}

\subsection{Discrete-time diagonal parametrization} \label{sec:ssl:discrete_diagonal}
One intuitive approach is that of parametrizing the SSL as a discrete-time complex-valued diagonal system, as proposed by \cite{orvieto2023resurrecting}.
\begin{subequations} \label{eq:ssl:parametrizations:discrete}
In particular,  $\tilde{A}$ can be parametrized as a Schur-stable diagonal matrix
\begin{equation}
    \tilde{A} = \tilde{\Lambda} = \diag(\lambda_1, ..., \lambda_{n_{\lambda}}).
\end{equation}
Each eigenvalue $j \in \{ 1, ..., n_\lambda \}$ is, in turn, parametrized by the modulus $- \exp(\mu_{j}) \in (0, 1)$ and the phase  $\exp(\theta_j)$, 
\begin{equation} \label{eq:ssl:parametrizations:discrete:eig}
    \lambda_j = \exp \Big(- \exp(\mu_{j})+ i \exp(\theta_j) \Big).
\end{equation}
\end{subequations}
{Note that since $\lvert \lambda_j \lvert < 1$ this parametrization guarantees that the SSM is structurally $\delta$ISS (Proposition~\ref{prop:diss})}. A relevant design choice advocated by \cite{orvieto2023resurrecting} is that of reparametrizing $\tilde{B}$ as
\begin{subequations} \label{eq:ssl:parametrizations:discrete_B}
\begin{equation}
    \tilde{B} = \diag(\gamma_1, ..., \gamma_{n_\lambda}) {\check{B}},
\end{equation}
where the normalization factor is defined as
\begin{equation}
    \gamma_j = \big( 1 - \lvert \lambda_j \lvert^2 \big)^{\frac{1}{2}}, \qquad \forall j \in \{ 1, ..., n_\lambda \}.
\end{equation}
\end{subequations}
This ensures that the rows of $\tilde{B}$ are normalized so that white noise inputs yield state trajectories with the same energy content as of the input.
Therefore, the set of learnable parameters of this SSL parametrization reads 
\begin{equation} \label{eq:ssl:parametrizations:discrete_theta}
    \Theta = \Big\{ \{ \mu_{j}, \theta_j \}_{j \in \{ 1, ..., n_{\lambda} \}}, \check{B}, \tilde{C}, D, F \Big\}.
\end{equation}

\subsection{Initialization strategy} \label{sec:parametrization:discrete:initialization}
\cite{orvieto2023resurrecting} propose to initialize the dynamics matrix randomly inside a portion of the circular crown lying within the unit circle. Letting $0 \leq \ubar{r} < \bar{r} < 1$ be the minimum and maximum modulus of each eigenvalue, respectively, and $0 \leq  \ubar{\theta} < \bar{\theta} < \pi$ be the minimum and maximum phase, 
\begin{equation} \label{eq:ssl:initialization:discrete_random}
\begin{aligned}
    \mu_j &\sim \mathfrak{U} \big[ \log(-\log(\bar{r})), \, \log(-\log(\ubar{r})) \big], \\
    \theta_j &\sim \mathfrak{U} \big[ \log(\log(\ubar{\theta})), \, \log(\log(\bar{\theta})) \big],
\end{aligned}
\end{equation}
where $\mathfrak{U}[a, b]$ denotes the uniform distribution with support $[a, b]$.
The complex matrices $\check{B}$ and $\tilde{C}$, and the real matrices $D$ and $F$\footnote{If $n_u = n_y$ the skip connection $F$ is usually fixed to the identity.} can be initialized with the Xavier initialization method \citep{kumar2017weight}. That is, they are sampled from a Normal distribution whose variance is scaled by a factor proportional to the number of columns.

\section{Continuous-time reparametrizations}
Another approach is that of parametrizing the SSL via a discretized continuous-time system, as proposed by \cite{gu2021efficiently} in the context of S4 models.
To this end, we let the SSL be parametrized as
\begin{equation} \label{eq:ssl:continuous}
    \begin{dcases}
        \dot{x}(t) =  A_c x(t) + B_c u(t), \\
        \eta(t) = C_c x(t), \\
        y(t) = \sigma(\eta(t)) + F u(t),
    \end{dcases}
\end{equation}
where --- similarly to \eqref{eq:ssl:conj} --- $\{ A_c, B_c, C_c \}$ are structured in terms of the blocks $\{ \tilde{A}_c, \tilde{B}_c, \tilde{C}_c \}$ 
\begin{equation} \label{eq:ssl:conj_cont}
\begin{aligned}
    A_c =& \begin{bmatrix}
        \Gamma \tilde{A}_c & \\ & \Gamma \conjop(\tilde{A}_c)
    \end{bmatrix}, \quad B_c = \begin{bmatrix}
        \Gamma \tilde{B}_c \\ \Gamma \conjop(\tilde{B}_c)
    \end{bmatrix}, \\
     C_c =& \begin{bmatrix}
            \tilde{C}_c & \conjop(\tilde{C}_c)
    \end{bmatrix},
\end{aligned}
\end{equation}
with $\tilde{A}_c \in \mathbb{C}^{n_\lambda \times n_\lambda}$, $\tilde{B}_c \in \mathbb{C}^{n_\lambda \times n_u}$,  $\tilde{C}_c \in \mathbb{C}^{n_y \times n_\lambda}$.
The factor $\Gamma$ in \eqref{eq:ssl:conj_cont} is a learnable timescale parameter, which can either be a real and positive scalar \citep{gu2021efficiently} or a real-valued, positive-definite diagonal matrix \citep{smith2022simplified}, i.e., $\Gamma = \diag(\gamma_1, ..., \gamma_{n_{\lambda}})$.

Under this parametrization, the discrete-time system matrices $\{ {A}, {B}, {C}, D \}$ can finally be expressed in terms of their continuous-time counterparts, $\{ A_c, B_c, C_c \}$, by defining the sampling time $\tau$ and a discretization method, such as forward Euler, the bilinear transform, or Zero-Order Hold (ZOH).
As shown in Section \ref{sec:ssl:discrete_diagonal}, an additional structure can now be imposed on $\tilde{A}_c$ to mitigate the overparametrization, thus making the training procedure more scalable, and to enforce the structural stability of the SSL {and hence, owing to Proposition~\ref{prop:diss}, the stability of the SSM}.
The possible structures of $\tilde{A}_c$ are now discussed.

\begin{subequations} \label{eq:ssl:parametrization:diagonal_continuous}
\subsection{Continuous-time diagonal parametrization}
One strategy, advocated by \cite{gu2022parameterization} and \cite{smith2022simplified}, is that of parametrizing $\tilde{A}_c$ as a diagonal matrix, i.e. $\tilde{A}_c = \tilde{\Lambda}_c$, where
\begin{equation}
    \tilde{\Lambda}_c = \diag \big( \lambda_1, ...,\lambda_{n_\lambda} \big).
\end{equation}    
Each eigenvalue $j \in \{ 1, ..., n_\lambda \}$ is, in turn, parametrized by the logarithm of its real and imaginary part, denoted as $\alpha_j^{\text{re}}$ and $\alpha_j^{\text{im}}$, respectively. That is,
\begin{equation}
    \lambda_j = -\exp(\alpha_j^{\text{re}}) + i \exp(\alpha_j^{\text{im}}).
\end{equation}
\end{subequations}
Note that, because $\mathfrak{Re}(\lambda_j) < 0$, this parametrization structurally guarantees the asymptotic stability of the SSL, {and hence the $\delta$ISS of the SSM (Proposition~\ref{prop:diss})}.
Overall, the set of learnable parameters is
\begin{equation} \label{eq:ssl:parametrization:diagonal_continuous_theta}
    \Theta = \Big\{ \{ \alpha_j^{\text{re}}, \alpha_j^{\text{im}}\}_{j \in \{1, ..., n_{\lambda} \}}, \tilde{B}_c, \tilde{C}_c, F, \Gamma \Big\}.
\end{equation}
As a side note, it is worth pointing out that the eigenvalues can also be parametrized in terms of the modulus and phase instead, as in \eqref{eq:ssl:parametrizations:discrete:eig}.

\subsection{Continuous-time DPLR parametrization} \label{sec:parametrization:continuous:dlpr}
The slightly more general Diagonal Plus Low-Rank (DPLR,  \cite{gu2021efficiently}) structure has also been proposed. 
According to this strategy, $\tilde{A}_c$ is parametrized as
\begin{equation} \label{eq:ssl:parametrization:dplr}
    \tilde{A}_c = \tilde{\Lambda}_c - \tilde{P} \tilde{Q}^*,
\end{equation}    
where $\tilde{\Lambda}_c$ is a diagonal component defined as
{
\begin{subequations}
\begin{equation}
    \tilde{\Lambda}_c = \diag(\lambda_1, ..., \lambda_{n_\lambda} )
\end{equation}
and $\tilde{P}, \tilde{Q} \in \mathbb{C}^{n_\lambda \times n_r}$ yield a low rank component, with $n_r \ll n_\lambda$.
In order to guarantee the structural stability of the SSL, and hence the $\delta$ISS of the SSM, one can take $\tilde{P} = \tilde{Q}$ (see \cite{gu2021efficiently}) and
\begin{equation}
    \lambda_j = -\phi(\alpha_{j}^{\text{re}}) + i \alpha_{j}^{\text{im}},
\end{equation}
with $\phi(\cdot)$ being any strictly-positive function, applied element-wise, e.g. $\phi(\alpha) = \max(0, \alpha) + \varepsilon$, with $\varepsilon > 0$. This ensures the negative definiteness of $\tilde{A}_c$.
\end{subequations}}
The set of learnable parameters, under the DPLR parametrization,~is
\begin{equation} \label{eq:ssl:parametrization:dplr_continuous_theta}
    \Theta = \Big\{ \{ \alpha_j^{\text{re}}, \alpha_j^{\text{im}}\}_{j \in \{1, ..., n_{\lambda} \}}, \tilde{P}, \tilde{Q}, \tilde{B}_c, \tilde{C}_c, F, \Gamma \Big\}.
\end{equation}

\subsection{Initialization strategies}
In the case of continuous-time DPLR parametrizations \cite{gu2021efficiently}  proposed to resort to the so-called HiPPO framework \citep{gu2020hippo} to initialize the learnable matrices $\tilde{B}_c$, $\tilde{\Lambda}_c$, $\tilde{P}$, and $\tilde{Q}$.
In particular, being it associated with long-term memory properties, the HiPPO matrix is regarded as a suitable initialization for \eqref{eq:ssl:parametrization:dplr}.
Such initialization is carried out by (\emph{i}) building the HiPPO-LegS matrix, (\emph{ii}) computing its normal and low-rank components, and (\emph{iii}) applying the eigenprojection, which yields the diagonal matrix $\tilde{\Lambda}_c$ and the projected low-rank components $\tilde{P}$ and $\tilde{Q}$. 
The HiPPO framework also provides an initialization for the input matrix $\tilde{B}_c$. 
For more details, the interested reader is referred to Appendix~\ref{appendix:hippo}.
The complex matrix $\tilde{C}_c$ and the real matrix $F$ are randomly initialized, e.g., with the Xavier method.

\cite{gu2022parameterization} proposed a similar strategy to initialize the learnable parameters of the continuous-time diagonal parametrizations~ \eqref{eq:ssl:parametrization:diagonal_continuous_theta}.
They suggested to initialize $\tilde{\Lambda}_c$ with the diagonal component of the normalized HiPPO-LegS matrix, discarding the low-rank terms $\tilde{P}$ and $\tilde{Q}$.

\begin{subequations} \label{eq:ssl:initialization:diagonal_continuous:modulus_phase}
As an alternative, we point out that a strategy similar to the one described in Section \ref{sec:parametrization:discrete:initialization} could be also applied to continuous-time diagonal parametrizations. 
Given a range for the modulus of eigenvalues, $0 < \ubar{r} < \bar{r} < \frac{\pi}{\tau}$, and for their phase, $\frac{\pi}{2} < \ubar{\theta} < \bar{\theta} \leq \pi$, one can sample
\begin{equation}
    \lambda_j = r \exp(i \theta), \quad r \sim \mathfrak{U}\big[ \ubar{r}, \bar{r}\big], \quad \theta \sim \mathfrak{U}\big[ \ubar{\theta}, \bar{\theta}\big],
\end{equation}
and initialize $\alpha_{j}^{\text{re}}$ and $\alpha_{j}^{\text{im}}$ as
\begin{equation}
    \alpha_{j}^{\text{re}} = \log \big( - \mathfrak{Re}(\lambda_j)\big), \quad \alpha_{j}^{\text{im}} = \log \big(\mathfrak{Im}(\lambda_j)\big).
\end{equation}
\end{subequations}
Under \eqref{eq:ssl:initialization:diagonal_continuous:modulus_phase}, $\Gamma$ can be initialized to the identity matrix.
The complex-valued matrices $\tilde{B}_c$ and $\tilde{C}_c$, and the real-valued matrix $F$ can be randomly initialized.

\subsection{Comments on the parametrization strategies} \label{sec:parametrization:continuous:problems}
For clarity purposes, the parameterizations have been introduced here in reverse chronological order.
The continuous-time DPLR parametrization of S4 has indeed been the one that sparked a renewed research interest in SSMs \citep{gu2021efficiently}. 
This strategy of parametrizing SSLs in the continuous-time domain has been proposed for two reasons.
Firstly, it enables the HiPPO framework to be used for the weights' initialization. 
Second, it allows simulation of the learned SSMs with a sampling time potentially different from that of the training data, while conventional (discrete-time) RNNs would call for a new training.

We point out, however, that the choice of parametrizing SSLs in the continuous-time domain implies the  \textit{aliasing} problem, usually overlooked in the SSMs literature.
Since~\eqref{eq:ssl:continuous} is discretized with the sampling time $\tau$ of the available training data, one should ensure that  (\emph{i}) the eigenvalues are not initialized beyond the Nyquist-Shannon bandwidth $\frac{\pi}{\tau}$ and (\emph{ii}) the eigenvalues remains within this bandwidth throughout the model's training. 
Eigenvalues beyond such bandwidth likely lead to reduced modeling capabilities and poor performances in extrapolation especially when different sampling times are adopted.
In this regard, we note that the HiPPO-based initialization strategies proposed by \cite{gu2021efficiently}, \cite{gu2022parameterization}, and \cite{smith2022simplified} do not, in general, meet these requirements. 
It is indeed well-known that the HiPPO matrix's eigenvalues grow quickly with $n_\lambda$.
To mitigate this issue, one should initialize the timescale parameter $\Gamma$ so that $\tilde{A}_c$ has eigenvalues within the Nyquist bandwidth.\footnote{In the SSM literature $\Gamma$ is often randomly sampled from a distribution which is uniform in its log-space \citep{smith2022simplified}.}

\section{Computational efficiency of SSM}
Because SSMs are learned with a simulation error minimization strategy, it is paramount to compute the output trajectory of the model --- given the input sequence $u_{0:T}$ --- as efficiently as possible.
As discussed in Remark \ref{remark:deep}, an SSM can be simulated by sequentially evaluating its SSLs.
In what follows, we therefore focus on how the generic SSL~\eqref{eq:ssl:discrete} can be efficiently simulated.

Since the SSL's dynamics are linear and asymptotically stable, and the static nonlinearity only affects the output variable, the state trajectory can be computed via the truncated Lagrange equation
\begin{equation} \label{eq:comp:conv_filter}
    x_{t+1} \approx \underbrace{\begin{bmatrix}
        A^{R-1} B & A^{R-2} B & ... & AB & B
    \end{bmatrix}}_{\mathfrak{B}_R} \cat(u_{t-R+1:t}),
\end{equation}
with the truncation $R \leq T$ being sufficiently large to ensure that the spectral radius of $A^R$ is negligible. 
In the SSM literature, \eqref{eq:comp:conv_filter} is known as the \emph{convolutional form}, as one can retrieve the state trajectory by convolving the input sequence $u_{0:T}$ with the \emph{filter} $\mathfrak{B}_R$,
\begin{subequations} \label{eq:comp:convolution}
\begin{equation} \label{eq:comp:convolution:state}
\begin{aligned}
    x_{1:T+1} =& \uncat \big( \mathfrak{B}_R \, \odot \cat( u_{0:T}) \big), \\
    \eta_t =& \, C x_t + D u_t, \qquad \forall t \in \{ 0, ..., T \}.
\end{aligned}
\end{equation}
The output $y_{0:T}$ can then be straightforwardly computed from $\eta_{0:T}$ by applying a static output transformation,
\begin{equation} \label{eq:comp:convolution:output}
    y_t = \sigma(\eta_t) + F u_t, \qquad \forall t \in \{ 0, ..., T \}.
\end{equation}
\end{subequations}
The convolutional form \eqref{eq:comp:convolution} enjoys a noteworthy computational efficiency since, as discussed in the remainder of this section, both the materialization of the filter $\mathfrak{B}_R$ and the convolution itself can be easily parallelized.
This allows the computational cost of simulating the SSL over a $T$-long input sequence to scale as $\mathcal{O}(T \log T)$ in place of the $\mathcal{O}(T^2)$ entailed by the iterative, sequential, application of~\eqref{eq:ssl:discrete} over the time axis \citep{smith2022simplified}.
The algorithms allowing such a cheap computation of \eqref{eq:comp:convolution:state} are now outlined.

\subsection{Convolution via Parallel Scan}
As proposed by \cite{smith2022simplified}, the task of scalably computing \eqref{eq:comp:convolution:state} 
can be addressed via Parallel Scan \citep{blelloch1990prefix}.
This approach stems from the idea that intermediate blocks of $\mathfrak{B}_R$ can be computed in parallel and combined, allowing to scale as $\mathcal{O}(T \log_2 T) $ times the complexity of the product $A \cdot A$.
While such a product in the particular case of diagonal parametrization scales as $\mathcal{O}(n_\lambda)$, for non-diagonal matrices its complexity is $\mathcal{O}(n_\lambda^3)$, thus rapidly increasing with the state dimension.
Thus, while the parallel scan approach can, in principle, be applied to any parametrization, it has primarily been utilized for diagonal ones, see \cite{orvieto2023resurrecting}.
For more details on how parallel scan is defined and implemented the reader is addressed to \cite{smith2022simplified}, Appendix~H.

\begin{rem}
    {Given the structure \eqref{eq:ssl:conj} of the SSL \eqref{eq:ssl:discrete}, one can instead simulate the subsystem  $\tilde{x}_{k+1} = \tilde{A} \tilde{x}_{k} + \tilde{B} {u}_{k}$ and $\tilde{\eta}_k = \tilde{C} \tilde{x}_k + D u_k$, and then compute the output as $y_k = \sigma(2 \, \mathfrak{Re}(\tilde{\eta}_k)) + F u_k$. This allows for further reduction of the computational burden without any approximations.}
\end{rem}

\subsection{Convolution via FFT}
For SSLs parametrized by the DPLR structure described in Section \ref{sec:parametrization:continuous:dlpr}, it is more efficient to carry out the convolution \eqref{eq:comp:convolution:state} in the frequency domain.
Operating in this domain, DPLR-parametrized SSLs can be simulated as efficiently as in diagonal parametrizations \citep{gu2021efficiently}.

Let us consider, for the purpose of explaination, the case $n_u = n_y = 1$.\footnote{This approach has been generalized to the case  $n_u > 1$ and/or $n_y > 1$ by applying it separately on each input-output pair.} 
Then, this approach consists in (\emph{i})~computing the Fast Fourier Transform (FFT) of the input signal $\mathcal{U}(\omega) = \FFT(u_{0:T})$, (\emph{ii}) multiplying it (frequency-wise) with the Fourier transform {$\tilde{\mathcal{H}}(\omega)$ of the discretized linear subsystem defined by the triplet $(\tilde{A}_c, \tilde{B}_c, \tilde{C}_c)$}, (\emph{iii}) computing the inverse FFT (iFFT), and (\emph{iv}) applying the nonlinear output transformation.
That is,
\begin{subequations}
\begin{align}
    \tilde{\eta}_{0:T} =& \, \iFFT \big( \tilde{\mathcal{H}}(\omega) \cdot \mathcal{U}(\omega) \big) , \label{eq:conv:fft} \\
    {\eta_t} =& {2 \, \mathfrak{Re}(\tilde{\eta}_t), \qquad\qquad  t \in \{ 0, ..., T \}},\\
    y_{t} =&\, \sigma({\eta}_{t}) + F u_{t}, \qquad\,  t \in \{ 0, ..., T \}.
\end{align}
\end{subequations}
Note that both the FFT and iFFT operations scale with complexity $\mathcal{O}(T \log T)$.
\cite{gu2021efficiently} have shown that, for continuous-time DPLR-parametrized SSLs (like S4), the Fourier transform {$\tilde{\mathcal{H}}(\omega)$} can be efficiently evaluated by resorting to black-box Cauchy kernels, see Appendix~\ref{appendix:fft}.
{With this approach, one issue to pay attention to is frequency leakage due to the input's nonperiodicity.}

\section{Numerical example} \label{sec:example}
To preliminary quantify the modeling capabilities of the described SSM architectures, we considered the well-known Silverbox benchmark \cite{wigren2013three}, for which a public dataset is available.
The Silverbox is an electronic circuit that mimics the input-output behavior of a damped mechanical system with a nonlinear elastic coefficient, and it has often been used for benchmarking, e.g., Wiener models \citep{tiels2015wiener}.
The training and validation data consist of ten experiments in which a multisine input with random frequency components is applied to the apparatus.
Each experiment features $8192$ input-output datapoints, collected with a sampling rate of $610.35$ Hz.
The validation dataset has been constructed by extracting $ 76$ subsequences (of length $T=512$ steps) from a random experiment.
From the remaining nine experiments $684$ training subsequences have been extracted.

The Silverbox dataset also contains an independent \textit{test dataset} that can be used to assess the accuracy of the identified models.
This test dataset consists of $40500$ samples collected by exciting the system with a filtered Gaussian noise having linearly increasing mean value.
As noted by \cite{tiels2015wiener}, the test dataset is characterized by two regions. 
The first $25000$ time-steps allow quantification of the model's accuracy in interpolation, i.e., within operating regions well explored by the training and validation data.
The successive time-steps allow, instead, enable the assessment of the model's extrapolation performances in operating regions not explored in the training and validation datasets, particularly in terms of the input signal amplitude.
For this reason, the modeling performances scored by the identified SSMs---measured by RMSE $[\text{mV}]$ and FIT index $[\%]$---are reported both on the overall test dataset and limitedly to the interpolatory regime.

\subsection{Identification results}
The training procedure was implemented in PyTorch~{2.1} and is described in more detail in the accompanying code\footnote{Source code: \url{https://github.com/bonassifabio/SSM-sysid}} or in Appendix \ref{appendix:training}. 
Our hope is that the code can help speed up continued research on these architectures.

The following SSM configurations have been considered for identification.
Additional details about their structures, initialization, and training hyperparameters are reported in Appendix~\ref{appendix:hyperparameters}.

\textbf{S4} \citep{gu2021efficiently} --- SSM whose layers are parametrized in the continuous-time domain \eqref{eq:ssl:continuous}-\eqref{eq:ssl:conj_cont} by a DLPR-structured state matrix \eqref{eq:ssl:parametrization:dplr} initialized via HiPPO.

\textbf{S5} \citep{smith2022simplified} --- SSM whose layers are parametrized by continuous-time diagonal systems \eqref{eq:ssl:continuous}-\eqref{eq:ssl:parametrization:diagonal_continuous} initialized via HiPPO.

\textbf{S5R} --- SSM whose layers are parametrized by continuous-time diagonal systems \eqref{eq:ssl:continuous}-\eqref{eq:ssl:parametrization:diagonal_continuous}. Similar to S5, but initialized by random sampling of the eigenvalues, \eqref{eq:ssl:initialization:diagonal_continuous:modulus_phase}.

\textbf{LRU} \citep{orvieto2023resurrecting} --- SSM whose layers are parametrized by discrete-time diagonal systems \eqref{eq:ssl:conj}-\eqref{eq:ssl:parametrizations:discrete_B}.

\begin{table}[t]
    \centering
    \caption{Performance of the identified SSMs}
    \label{tab:model_comparison}
    \resizebox{\columnwidth}{!}{
    \begin{tabular}{lccccc}
        \toprule
        {Model} & \multicolumn{2}{c}{{First 25000 steps}} & \multicolumn{2}{c}{{Full}} \\
        \cmidrule(lr){2-3} \cmidrule(lr){4-5}
        & {RMSE} & {FIT} & {RMSE} & {FIT} \\
        \midrule
        S4 ($L=4$, $n_\lambda=10$)\textsuperscript{\ref{table_note:nyquist}} & $0.81$ & $97.60$ & $4.73$ &	$95.49$ \\       
        S5 ($L=4$, $n_\lambda=10$)\tablefootnote{\label{table_note:nyquist}At least one eigenvalue falls beyond the Nyquist frequency.} & $0.73$& $97.78$	& $3.56$ & $96.48$  \\
        S5R ($L=4$, $n_\lambda=10$) & $0.77$& $97.62$	& $4.58$ & $95.92$  \\
        LRU ($L=4$, $n_\lambda=10$) & $0.73$ & $97.71$ & $4.18$ & $96.37$ \\
        \midrule
        TCNN \citep{andersson2019deep} & 0.75 & - & 4.9 & - \\
        LSTM \citep{andersson2019deep} & 0.31 & - & 4.0 & - \\
        BLA \citep{tiels2015wiener} & - & - & 13.7 & - \\
        Wiener \citep{tiels2015wiener}& 1.9 & - & 9.2 & - \\
        Grey-box NARX\citep{ljung2004estimation} & - & - & 0.3 & - \\
        \bottomrule \\
    \end{tabular}}
\end{table}

In Table \ref{tab:model_comparison} the performance metrics scored by the best identified SSM are reported, and they are compared to some of those reported in the literature\footnote{For detailed accounts of the literature, the reader is referred to \cite{andersson2019deep} and \cite{maroli2019nonlinear}.}, while in Appendix~\ref{appendix:hyperparameters} the trajectory of the simulation error is reported.
Note that, although these SSMs are more accurate than traditional Wiener models, they are still in line with those achieved  by other deep learning models like TCNNs and LSTMs{, while featuring significantly less learnable parameters\footnote{A diagonal SSL parametrized in continuous-time has $n_\lambda (3 + 2 n_y + 2 n_u) + n_y n_u$ learnable parameters, compared to the $n_x(4 + 3n_x + n_u+n_y)+n_y$ learnable parameters of an LSTM layer.} and high parallelizability of the training procedure}.

At last, let us point out that training competitive SSMs requires, in general, a careful selection of the architecture hyperparameters.
These not only include the number of layers and the state and output size of each layer, but also a suitable initialization strategy for the selected parametrization.
The computational efficiency of these models and their reduced parameter footprint come at the cost of a slightly increased architecture design effort.
Further testing of SSMs for nonlinear system identification is thus advisable to establish empirical design criteria.

\section{Conclusions and research directions}
Structured State-space Models represent an interesting approach to identifying deep Wiener models. 
In this paper we summarized recent developments of SSMs in the machine learning community, attempting to disentangle their structure, parameterization, initialization, and simulation~aspects.
The goal was to make these models more accessible for the system identification community, to stimulate further investigation of these models for identification.
We believe there is scope for interesting future developments---possible research directions are summarized below.

\emph{Minimal structures} \quad Structures with fewer learnable parameters should be considered, as long as they ensure the model's stability and allow for an efficient simulation. 
{In this context, it would be advisable to evaluate whether parameterizations in the transfer function domain, similar to dynoNet \cite{forgione2021dynonet}, yield less overparametrized and more effective models.}

\emph{Data-driven initializations} \quad Novel  strategies could exploit training data to improve and tailor the initialization of learnable parameters. For Wiener models, for example, initialization based on best linear approximators are known to boost the identified model performances.

\emph{Discrete- or  continuous-time parametrizations?} \quad Despite the revived interest in SSMs originated from continuous-time parameterizations, \cite{orvieto2023resurrecting} questioned their real need, showing how similar performances can be achieved also with discrete-time parameterizations. 
This issue should also be further investigated in light of the aliasing problem {discussed in Section \ref{sec:parametrization:continuous:problems}}.

\textit{Extensive benchmarking} \quad The performances of SSMs for nonlinear system identification should be assessed on a broader set of benchmarks, possibly featuring long-term time dependencies, for which SSMs were proposed.

\bibliography{bibliography}

\begin{thebibliography}{27}
\providecommand{\natexlab}[1]{#1}
\providecommand{\url}[1]{\texttt{#1}}
\providecommand{\urlprefix}{URL }
\expandafter\ifx\csname urlstyle\endcsname\relax
  \providecommand{\doi}[1]{doi:\discretionary{}{}{}#1}\else
  \providecommand{\doi}{doi:\discretionary{}{}{}\begingroup \urlstyle{rm}\Url}\fi

\bibitem[{Andersson et~al.(2019)Andersson, Ribeiro, Tiels, Wahlstr{\"o}m, and Sch{\"o}n}]{andersson2019deep}
Andersson, C., Ribeiro, A.H., Tiels, K., Wahlstr{\"o}m, N., and Sch{\"o}n, T.B. (2019).
\newblock Deep convolutional networks in system identification.
\newblock In \emph{2019 IEEE 58th Conference on Decision and Control (CDC)}, 3670--3676. IEEE.

\bibitem[{Angeli(2002)}]{angeli2002lyapunov}
Angeli, D. (2002).
\newblock A lyapunov approach to incremental stability properties.
\newblock \emph{IEEE Transactions on Automatic Control}, 47(3), 410--421.

\bibitem[{Bengio et~al.(2017)Bengio, Goodfellow, and Courville}]{goodfellow2016deep}
Bengio, Y., Goodfellow, I., and Courville, A. (2017).
\newblock \emph{Deep learning}, volume~1.
\newblock MIT press Massachusetts, USA.

\bibitem[{Bianchi et~al.(2017)Bianchi, Maiorino, Kampffmeyer, Rizzi, and Jenssen}]{bianchi2017recurrent}
Bianchi, F.M., Maiorino, E., Kampffmeyer, M.C., Rizzi, A., and Jenssen, R. (2017).
\newblock \emph{Recurrent neural networks for short-term load forecasting: an overview and comparative analysis}.
\newblock Springer.

\bibitem[{Blelloch(1990)}]{blelloch1990prefix}
Blelloch, G.E. (1990).
\newblock Prefix sums and their applications.
\newblock In J.H. Reif (ed.), \emph{Synthesis of parallel algorithms}. Morgan Kaufmann Publishers Inc.

\bibitem[{Bonassi et~al.(2022)Bonassi, Farina, Xie, and Scattolini}]{bonassi2022survey}
Bonassi, F., Farina, M., Xie, J., and Scattolini, R. (2022).
\newblock {O}n {R}ecurrent {N}eural {N}etworks for learning-based control: recent results and ideas for future developments.
\newblock \emph{Journal of Process Control}, 114, 92--104.

\bibitem[{Bonassi et~al.(2024)Bonassi, {La Bella}, Farina, and Scattolini}]{bonassi2024nonlinear}
Bonassi, F., {La Bella}, A., Farina, M., and Scattolini, R. (2024).
\newblock Nonlinear {MPC} design for incrementally {ISS} systems with application to {GRU} networks.
\newblock \emph{Automatica}, 159, 111381.

\bibitem[{Forgione and Piga(2021)}]{forgione2021dynonet}
Forgione, M. and Piga, D. (2021).
\newblock {dynoNet}: A neural network architecture for learning dynamical systems.
\newblock \emph{International Journal of Adaptive Control and Signal Processing}, 35(4), 612--626.

\bibitem[{Gu et~al.(2020)Gu, Dao, Ermon, Rudra, and R{\'e}}]{gu2020hippo}
Gu, A., Dao, T., Ermon, S., Rudra, A., and R{\'e}, C. (2020).
\newblock Hippo: Recurrent memory with optimal polynomial projections.
\newblock \emph{Advances in neural information processing systems}, 33, 1474--1487.

\bibitem[{Gu et~al.(2022)Gu, Goel, Gupta, and R{\'e}}]{gu2022parameterization}
Gu, A., Goel, K., Gupta, A., and R{\'e}, C. (2022).
\newblock On the parameterization and initialization of diagonal state space models.
\newblock \emph{Advances in Neural Information Processing Systems}, 35, 35971--35983.

\bibitem[{Gu et~al.(2021)Gu, Goel, and R{\'e}}]{gu2021efficiently}
Gu, A., Goel, K., and R{\'e}, C. (2021).
\newblock Efficiently modeling long sequences with structured state spaces.
\newblock \emph{arXiv preprint arXiv:2111.00396}.

\bibitem[{Gupta et~al.(2022)Gupta, Gu, and Berant}]{gupta2022diagonal}
Gupta, A., Gu, A., and Berant, J. (2022).
\newblock Diagonal state spaces are as effective as structured state spaces.
\newblock \emph{Advances in Neural Information Processing Systems}, 35, 22982--22994.

\bibitem[{Kumar(2017)}]{kumar2017weight}
Kumar, S.K. (2017).
\newblock On weight initialization in deep neural networks.
\newblock \emph{arXiv preprint arXiv:1704.08863}.

\bibitem[{Lanzetti et~al.(2019)}]{lanzetti2019recurrent}
Lanzetti, N. et~al. (2019).
\newblock Recurrent neural network based {MPC} for process industries.
\newblock In \emph{2019 18th European Control Conference (ECC)}, 1005--1010. IEEE.

\bibitem[{Ljung et~al.(2004)Ljung, Zhang, Lindskog, and Juditski}]{ljung2004estimation}
Ljung, L., Zhang, Q., Lindskog, P., and Juditski, A. (2004).
\newblock Estimation of grey box and black box models for non-linear circuit data.
\newblock \emph{IFAC Proceedings Volumes}, 37(13), 399--404.

\bibitem[{Marconato et~al.(2013)Marconato, Sj{\"o}berg, Suykens, and Schoukens}]{marconato2013improved}
Marconato, A., Sj{\"o}berg, J., Suykens, J.A., and Schoukens, J. (2013).
\newblock Improved initialization for nonlinear state-space modeling.
\newblock \emph{IEEE Transactions on instrumentation and Measurement}, 63(4), 972--980.

\bibitem[{Maroli et~al.(2019)Maroli, {\"O}zg{\"u}ner, and Redmill}]{maroli2019nonlinear}
Maroli, J.M., {\"O}zg{\"u}ner, {\"U}., and Redmill, K. (2019).
\newblock Nonlinear system identification using temporal convolutional networks: a silverbox study.
\newblock \emph{IFAC-PapersOnLine}, 52(29), 186--191.

\bibitem[{Miller and Hardt(2019)}]{miller2018stable}
Miller, J. and Hardt, M. (2019).
\newblock Stable recurrent models.
\newblock In \emph{International Conference on Learning Representations}.
\newblock ArXiv preprint arXiv:1805.10369.

\bibitem[{Orvieto et~al.(2023)}]{orvieto2023resurrecting}
Orvieto, A. et~al. (2023).
\newblock Resurrecting recurrent neural networks for long sequences.
\newblock \emph{arXiv preprint arXiv:2303.06349}.

\bibitem[{Ramachandran et~al.(2017)Ramachandran, Zoph, and Le~Quoc}]{ramachandran2017searching}
Ramachandran, P., Zoph, B., and Le~Quoc, V. (2017).
\newblock Searching for activation functions.
\newblock \emph{arXiv preprint arXiv:1710.05941}.

\bibitem[{Schoukens and Tiels(2017)}]{schoukens2017identification}
Schoukens, M. and Tiels, K. (2017).
\newblock Identification of block-oriented nonlinear systems starting from linear approximations: A survey.
\newblock \emph{Automatica}, 85, 272--292.

\bibitem[{Smith et~al.(2022)Smith, Warrington, and Linderman}]{smith2022simplified}
Smith, J.T., Warrington, A., and Linderman, S. (2022).
\newblock Simplified state space layers for sequence modeling.
\newblock In \emph{The Eleventh International Conference on Learning Representations}.

\bibitem[{Sun and Wei(2022)}]{sun2022efficient}
Sun, Y. and Wei, H.L. (2022).
\newblock Efficient mask attention-based narmax (mab-narmax) model identification.
\newblock In \emph{2022 27th International Conference on Automation and Computing (ICAC)}, 1--6. IEEE.

\bibitem[{Tiels(2015)}]{tiels2015wiener}
Tiels, K. (2015).
\newblock \emph{Wiener system identification with generalized orthonormal basis functions}.
\newblock Ph{D} thesis, Vrije Universiteit Brussell.

\bibitem[{Wigren and Schoukens(2013)}]{wigren2013three}
Wigren, T. and Schoukens, J. (2013).
\newblock Three free data sets for development and benchmarking in nonlinear system identification.
\newblock In \emph{2013 European control conference (ECC)}, 2933--2938. IEEE.

\bibitem[{Wills and Ninness(2012)}]{wills2012generalised}
Wills, A. and Ninness, B. (2012).
\newblock Generalised hammerstein--wiener system estimation and a benchmark application.
\newblock \emph{Control Engineering Practice}, 20(11), 1097--1108.

\bibitem[{Yu et~al.(2018)Yu, Ljung, and Verhaegen}]{yu2018identification}
Yu, C., Ljung, L., and Verhaegen, M. (2018).
\newblock Identification of structured state-space models.
\newblock \emph{Automatica}, 90, 54--61.

\end{thebibliography}

\appendix

\section{HiPPO-LegS Matrix} \label{appendix:hippo}
In this appendix we provide more information on the construction of the HiPPO-LegS matrix, and how this matrix is used to initialize the $\tilde{\Lambda}_c$, $\tilde{P}$, and $\tilde{Q}$ matrices of the DPLR parameterization \eqref{eq:ssl:parametrization:dplr}.
The HiPPO framework is built on the idea of projecting the data into a high-dimensional latent space characterized by an orthogonal polynomial basis \cite{gu2020hippo}. 
Consistently with the Koopman theory, within this latent state any dynamical system can be represented by an infinite-dimensional LTI system.
The HiPPO framework gained popularity since both the projection operator and the latent linear representation admit a simple closed-form expression.

\begin{subequations}
Depending on the metric used to define the projection operator, different HiPPO matrices can be defined --- we here focus on the HiPPO-LegS matrix, constructed by taking a scaled Legendre measure.
This matrix has the form 
\begin{equation}
    A_{\textrm{LegS}} = {A}_{\textrm{LegS}}^{(N)} - p p^\prime,
\end{equation}
where {${A}_{\textrm{LegS}}^{(N)} \in \mathbb{R}^{n_{\lambda} \times n_{\lambda}}$} is the orthonormal component, and $p \in {\mathbb{R}^{n_{\lambda}}}$ some $1$-rank component. 
Denoting by $\big[{A}_{\textrm{LegS}}^{(N)} \big]_{a,b}$ the element of ${A}_{\textrm{LegS}}^{(N)}$ at position $(a, b)$, it holds that
\begin{equation}
    \big[{A}_{\textrm{LegS}}^{(N)} \big]_{a,b} = - \begin{dcases}
    -\Big( a - \frac{1}{2} \Big)^{\frac{1}{2}}  \cdot \Big( b - \frac{1}{2} \Big)^{\frac{1}{2}} & a < b, \\
    \frac{1}{2} & a = b, \\
    \Big( a - \frac{1}{2} \Big)^{\frac{1}{2}}  \cdot \Big( b - \frac{1}{2} \Big)^{\frac{1}{2}} & a > b.
\end{dcases}
\end{equation}
The low-rank component $p$ is instead defined as
\begin{equation}
    \big[ p \big]_a = \Big( a - \frac{1}{2} \Big)^{\frac{1}{2}}.
\end{equation}
\end{subequations}

Letting the normal component be eigendecomposed as ${A}_{\textrm{LegS}}^{(N)} = V \Lambda_c V^*$, the HiPPO-LegS matrix is projected as 
\begin{equation} \label{eq:appendix:hippo_dplr}
    V^* {A}_{\textrm{LegS}} V = {\tilde{\Lambda}_c - \underbrace{V^* p}_{\tilde{P}} \underbrace{p^{\prime} V}_{\tilde{Q}^*} := \tilde{A}_c}.
\end{equation}

The $\tilde{B}_c$ matrix is also given by the HiPPO framework as
\begin{equation}\label{eq:appendix:hippo_b}
    \big[\tilde{B}_c \big]_{a, b} = \sqrt{2(a-1) +1}.
\end{equation}

\section{FFT-based convolution for DPLR parametrizations} \label{appendix:fft}

Consider the transfer function $\tilde{H}(s) = \tilde{C}_c (sI - \tilde{A}_c)^{-1} \tilde{B}_c$, where the state matrix $\tilde{A}_c$ \eqref{eq:ssl:conj_cont} is structured according to the DPLR parametrization \eqref{eq:ssl:parametrization:dplr} and, without loss of generality, $\Gamma = 1$.
{\cite{gu2021efficiently} propose to compute $\tilde{\eta}_{0:T}$ as the response of $\tilde{H}(s)$, from which $\eta$ can then be computed as $\eta_t = 2 \mathfrak{Re} \big(\tilde{\eta}_{t}\big)$ for any $t \in \{0, ..., T \}$.}

{To retrieve $\tilde{\eta}_{0:T}$ efficiently, $\tilde{H}(s)$ is first discretized via the bilinear transformation $s = \frac{2}{\tau} \frac{1 - z^{-1}}{1 + z^{-1}}$, which yields
\begin{equation} \label{eq:appendix:tf_z}
    \tilde{H}(z) = \tilde{C}_c \left( \frac{2}{\tau} \frac{1 - z^{-1}}{1 + z^{-1}} I - \tilde{A}_c \right)^{-1} \tilde{B}_c.
\end{equation}
Computing \eqref{eq:appendix:tf_z} involves inverting an $n_{\lambda} \times n_{\lambda}$ matrix which might be impractical.
\cite{gu2021efficiently} noted that since $\tilde{A}_c = \tilde{\Lambda}_c - \tilde{P} \tilde{Q}^*$, by applying the Matrix Inversion Lemma \eqref{eq:appendix:tf_z} can be re-worked as 
\begin{subequations} \label{eq:appendix:matrix_inv_1}
\begin{equation}
\scalemath{0.9}{
    \tilde{H}(z) = \tilde{C}_c \left[ \tilde{W}(z)  - \tilde{W}(z) \tilde{P} ( I + \tilde{Q}^* \tilde{W}(z) \tilde{P})^{-1} \tilde{Q}^* \tilde{W}(z) \right]^{-1} \tilde{B}_c,}
\end{equation}
where 
\begin{equation}
    \tilde{W}(z) = \left( \frac{2}{\tau} \frac{1 - z^{-1}}{1 + z^{-1}} I - \tilde{\Lambda}_c \right)^{-1}.
\end{equation}
\end{subequations}}
Note that \eqref{eq:appendix:matrix_inv_1} involves the inversion of an $n_r \times n_r$ matrix, where the low-rank dimension $n_r$ is often $1$ --- as in Appendix \ref{appendix:hippo}. 
Moreover, $\tilde{W}(z)$ can be easily computed since $\tilde{\Lambda}_c$ is diagonal.
Finally, the Fourier transform is retrieved evaluating the transfer function on the unit circle,
\begin{equation} \label{eq:appendix:frf}
    {\tilde{\mathcal{H}}(\omega)} = \left. {\tilde{H}(z)} \right\lvert_{z = \exp(i \omega \tau)},
\end{equation}
for $\omega \in \{ \omega_0, ..., \omega_{T-1} \}$, where $\omega_r = \frac{2 \pi}{\tau} \frac{r}{T}$.
As shown by \cite{gu2021efficiently},  \eqref{eq:appendix:matrix_inv_1} and \eqref{eq:appendix:frf} can be reconducted to a black-box Cauchy kernel, and can be easily (and efficiently) implemented even for high-dimensional systems.

\section{Training procedure} \label{appendix:training}
Like most RNN architectures, SSMs are learned via simulation error minimization, see \cite{bianchi2017recurrent}. 
To this end, assume that a collection of input-output sequences are collected from the plant to be identified via a suitably-designed experiment campaign. 
We let such a dataset be denoted as $\mathcal{D} = \big\{ \big(u_{0:T}^{\{l\}}, y_{0:T}^{\{l\}} \big)_{l \in \mathcal{I}} \big\}$, where the superscript $^{\{l\}}$ is used to index the sequences over the set $\mathcal{I} = \{ 1, ..., N \}$.
For compactness, it is assumed that these input-output sequences have a fixed length $T$, and that the input-ouput data have been normalized \citep{goodfellow2016deep}.
The sequences are partitioned into a training set $\mathcal{I}_{\text{tr}} \subset \mathcal{I}$ and a validation set $\mathcal{I}_{\text{val}} \subset \mathcal{I}$, where $\mathcal{I}_{\text{tr}} \cup \mathcal{I}_{\text{val}} = \mathcal{I}$ and $\mathcal{I}_{\text{tr}} \cap \mathcal{I}_{\text{val}} = \emptyset$.

The training procedure is carried out iteratively. 
At every such iteration (\emph{epoch}), the training set is randomly split into $M$ independent partitions (\emph{batches}), denoted as  $\mathcal{I}^{\{ m \}}$, with $m \in \{ 1, ..., M\}$.
For each of these batches, the loss function is defined as the average simulation Mean Squared Error (MSE), i.e.
\begin{equation} \label{eq:training:loss}
    \mathcal{L}(\mathcal{I}; \Theta) = \frac{1}{\lvert \mathcal{I} \lvert} \sum_{l \in \mathcal{I}} \MSE \Big( y_{0:T}\big(u_{0:T}^{\{l\}}, \Theta \big),  y_{0:T}^{\{l\}} \Big),
\end{equation}
where $y_{0:T}\big(u_{0:T}^{\{l\}}, \Theta \big)$ denotes the free-run simulation of the SSM \eqref{eq:ssl:discrete}, computed applying  \eqref{eq:comp:convolution} sequentially over the $L$ SSLs.
The parameters are then updated via a gradient descent step.
In the simplest case of first-order gradient methods, this update reads 
\begin{equation}\label{eq:training:backprop}
    \Theta \gets \Theta - \rho \nabla_{\Theta} \mathcal{L}\big( \mathcal{I}^{\{m\}}; \Theta\big),
\end{equation}
where $\nabla_{\Theta}$ denotes the gradient operator and $\rho > 0$ is the learning rate.

The training procedure is generally repeated for a prescribed number of epochs, or until some validation metrics, e.g. $\mathcal{L}(\mathcal{I}_{\text{val}}; \Theta)$, stops improving \citep{bianchi2017recurrent}.
The resulting training procedure is summarized in Algorithm~\ref{alg:modeltraining}.

\begin{algorithm}[t]
\caption{Training procedure}\label{alg:modeltraining}
\begin{algorithmic}
\State Selection of the SSM structure and parameterization
\State Initialization of the learnable parameters $\Theta$
\For{epoch $e \in \{ 0, ..., E \}$} 
	\State \text{Partition $\mathcal{I}_{\text{tr}}$ into random batches $\mathcal{I}^{\{1\}}, ..., \mathcal{I}^{\{M\}}$}
	\For{\text{batch} $m \in \{ 1, ..., M \}$} 
		\State Compute the loss $\mathcal{L}(\mathcal{I}^{\{m\}}; \Theta)$ via \eqref{eq:training:loss}
        \State Backpropagation e.g. via \eqref{eq:training:backprop} 
	\EndFor
	\State Evaluate validation loss $\mathcal{L}(\mathcal{I}_{\text{val}}; \Theta)$ for early stopping
\EndFor
\end{algorithmic}
\end{algorithm}

\section{Hyperparameters and performances of the trained SSMs}  \label{appendix:hyperparameters}
In this appendix details on the structures, initializations, and training hyperparameters of the SSMs described in Section \ref{sec:example} are reported.
Note that, for all these architectures, (\emph{i}) the ELU activation function \citep{ramachandran2017searching} was used for $\sigma(\cdot)$, and (\emph{ii}) the output size of intermediate layers ($\ell \in \{ 1, ..., L -1\}$) was fixed to $4$.

\textbf{S4} --- The adopted S4 model consists of $L=4$ SSLs, each parametrized in continuous-time with $n_\lambda = 10$ learnable eigenvalues.
The HiPPO-LegS matrix \eqref{eq:appendix:hippo_dplr} was used to initialize the DLPR state matrix \eqref{eq:ssl:parametrization:dplr}, and \eqref{eq:appendix:hippo_b} was used to initialize the input matrix $\tilde{B}_c$.
The timescale parameter $\Gamma$ was initialized randomly in the range $(20, 100)$.
The output matrix $\tilde{C}_c $ was initialized with the Xavier initialization \citep{kumar2017weight}.

\textbf{S5} --- The adopted S5 model features $L=4$ layers with $n_\lambda = 10$ learnable eigenvalues parametrized in continuous-time.
The diagonal state matrix $\tilde{\Lambda}_c$ was initialized to the diagonalized normal component of the HiPPO-LegS matrix, while \eqref{eq:appendix:hippo_b} was used to initialize the input matrix $\tilde{B}_c$.
The scalar $\Gamma$ was initialized to $25$, whereas $\tilde{C}_c $ was initialized with the Xavier  method.

\textbf{S5R} --- The S5R model has $L=4$ layers with $n_\lambda = 10$ learnable eigenvalues parametrized in continuous-time.
For the initialization, the eigenvalues of the diagonal matrix $\tilde{\Lambda}_c$ were randomly sampled using \eqref{eq:ssl:initialization:diagonal_continuous:modulus_phase}, with $(\ubar{r}, \bar{r}) = \big(0.1 \frac{\pi}{\tau}, \frac{\pi}{\tau} \big)$ and $(\ubar{\theta}, \bar{\theta}) = \big(\frac{\pi}{6}, \frac{3\pi}{4} \big)$.
The scalar $\Gamma$ was initialized to $0.9$, whereas  $\tilde{B}_c$ and $\tilde{C}_c $ were initialized with the Xavier method.

\textbf{LRU} --- The adopted LRU has $L=4$ layers with $n_\lambda = 10$ learnable eigenvalues. This model is parametrized in the discrete-time domain. 
The diagonal state matrix was initialized via \eqref{eq:ssl:initialization:discrete_random} by randomly sampling its eigenvalues from the circular crown sector delimited by $(\ubar{r}, \bar{r}) = (0.05, 0.975) \subset (0, 1)$ and $(\ubar{\theta}, \bar{\theta}) = (0, 2\pi)$.
The Xavier initialization method was used for matrices $\tilde{\tilde{B}}$ and $\tilde{C}$, while $D$ was initialized to the null matrix.

\smallskip
The training procedures were conducted with the Adam optimizer \citep{goodfellow2016deep}, with a batch size of $40$ and an initial learning rate $\rho = 0.003$.
A learning rate scheduler was  used to dynamically adjust $\rho$ on plateaus, reducing it by $20\%$ after $30$ epochs without improvements on the training data.
An early stopping procedure was included to halt the training after $150$ epochs without improvement on the validation metrics, and still within at most $2750$ epochs.

\begin{figure}[t]
    \centering
    \subfloat[S4]{
    \includegraphics[width=0.5\columnwidth]{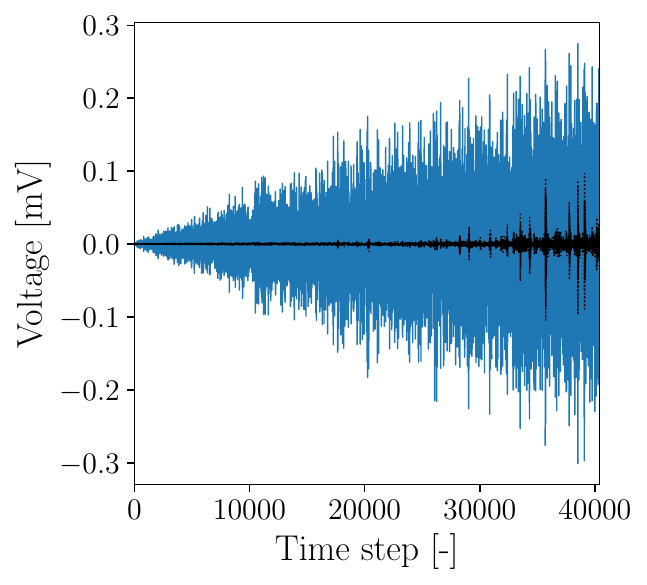}}  
    \subfloat[S5]{
    \includegraphics[width=0.5\columnwidth]{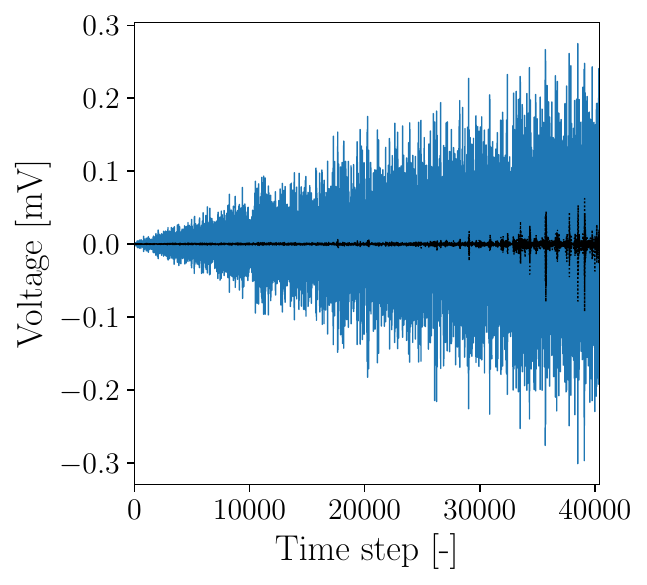}}  \\
    \subfloat[S5R]{
    \includegraphics[width=0.5\columnwidth]{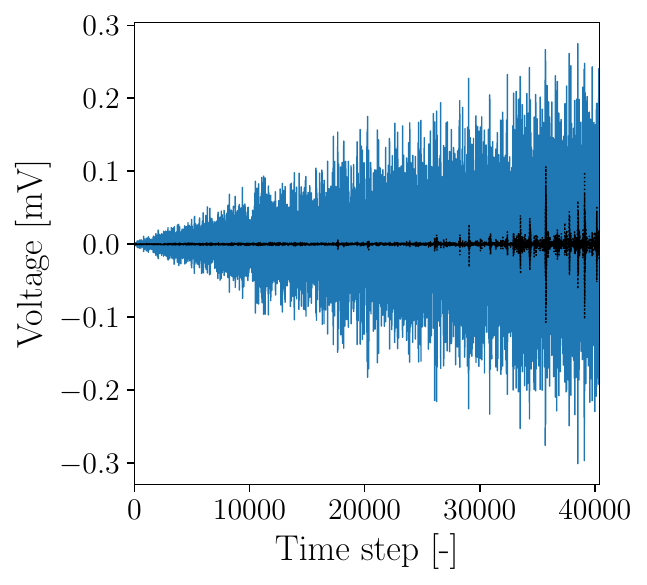}} 
    \subfloat[LRU]{
    \includegraphics[width=0.5\columnwidth]{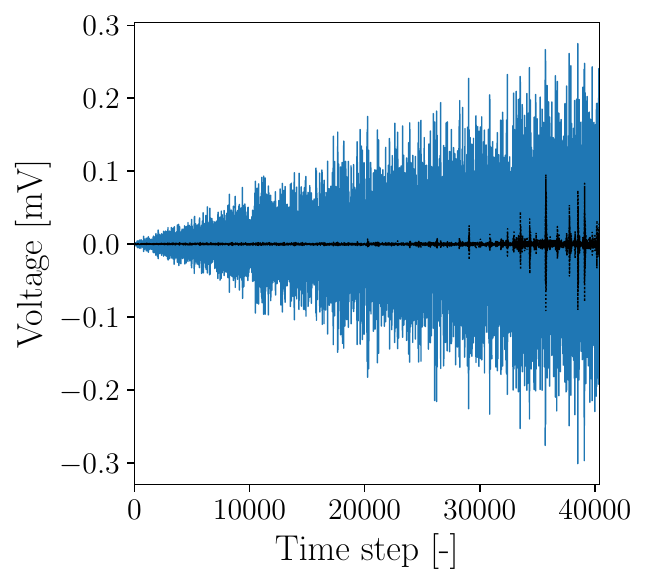}}
    \caption{Free-run simulation error of the trained SSMs (black dotted line) with respect to the ground truth (blue line) over the entire test dataset.}
    \label{fig:test}
\end{figure}

In Figure \ref{fig:test} the free-run simulation error on the test dataset of each SSM's best training instance is depicted.
As expected, this error is fairly limited in the first $25000$ time-steps of the test dataset, where the model operates in a region well explored by the training data,  while it is larger in the second part of the dataset, where the model operates in extrapolation \citep{andersson2019deep}.

\end{document}